\documentclass[12pt,a4paper,aps,preprint,superscriptaddress,nofootinbib]{revtex4-1}
\usepackage[utf8]{inputenc}
\usepackage{graphicx}
\usepackage{cancel}
\usepackage{amssymb}
\usepackage{textcomp}
\usepackage{amsmath}
\usepackage{bm}
\usepackage{times}
\usepackage{epsfig}
\usepackage{epstopdf}
\usepackage{color}
\usepackage{multirow}
\usepackage[colorlinks=true, pdfstartview=FitV, linkcolor=blue, citecolor=blue, urlcolor=blue]{hyperref}
\usepackage{ulem}

\def\lsim{\mathrel{\raise.3ex\hbox{$<$\kern-.75em\lower1ex\hbox{$\sim$}}}}
\def\gsim{\mathrel{\raise.3ex\hbox{$>$\kern-.75em\lower1ex\hbox{$\sim$}}}}

\definecolor{orange}{rgb}{1,0.5,0}

\def\beq{\begin{equation}}
\def\eeq{\end{equation}}
\def\be{\begin{equation}}
\def\ee{\end{equation}}
\def\bea{\begin{eqnarray}}
\def\eea{\end{eqnarray}}

\def\ev{\,{\rm eV}}
\def\to{\rightarrow}

\newcommand{\minigraph}[5][0.25in]{\begin{minipage}{#2}\begin{center}\includegraphics[width=#2]{#5}\\\vspace{#3}\hspace{#1}{\footnotesize #4}\end{center}\end{minipage}}

\begin{document}

\title{Searching for heavy neutrino in terms of tau lepton at future hadron collider}

\author{Chengcheng Han}
\email{hanchch@mail.sysu.edu.cn}
\affiliation{
School of Physics, Sun Yat-Sen University, Guangzhou 510275, China
}

\author{Tong Li}
\email{litong@nankai.edu.cn}
\affiliation{
School of Physics, Nankai University, Tianjin 300071, China
}

\author{Chang-Yuan Yao}
\email{yaocy@nankai.edu.cn}
\affiliation{
School of Physics, Nankai University, Tianjin 300071, China
}

\begin{abstract}
The tau lepton plays important role in the correlation between the low-energy neutrino oscillation data and the lepton flavor structure in heavy neutrino decay. We investigate the lepton flavor signatures with tau lepton at hadron collider through lepton number violating (LNV) processes. In the Type I Seesaw with U$(1)_{\rm B-L}$ extension, we study the pair production of heavy neutrinos via a $Z'$ resonance. We present a detailed assessment of the search sensitivity to the channels with tau lepton in the subsequent decay of heavy neutrinos. For the benchmark model with $Z'$ only coupled to the third generation fermions, we find that the future circular collider (FCC-hh) can discover the LNV signal with tau lepton for $M_{Z'}$ up to 2.2 (3) TeV with the gauge coupling $g'=0.6$ and the integrated luminosity of 3 (30) ab$^{-1}$. The test on the flavor combinations of SM charged leptons would reveal the specific nature of different heavy neutrinos.
\end{abstract}

\maketitle

\section{Introduction}

It is well-known that, in the context of the Standard Model (SM), the non-zero but small Majorana neutrino masses can be realized at leading order through a dimension-five operator~\cite{Weinberg:1979sa}
\begin{eqnarray}
{\kappa\over \Lambda}l_Ll_LHH,
\label{weinberg}
\end{eqnarray}
where $l_L$ and $H$ stand for the SM lepton doublet and the Higgs doublet, respectively.
To receive the above ``Weinberg operator'', the minimal extension of the SM content permits only the Type I, Type II or Type III Seesaw mechanism at tree level~\cite{Ma:1998dn}. In particular, in Type I Seesaw model~\cite{Minkowski:1977sc,Yanagida:1979as,GellMann:1980vs,Glashow:1979nm,Mohapatra:1979ia,Shrock:1980ct,Schechter:1980gr,Davidson:1987mh}, the neutrino mass matrix is generated by at least two generations of fermionic multiplets as SU$(2)_L$ singlet.
The new physics scale $\Lambda$ in Eq.~(\ref{weinberg}) is replaced by the mass of the new multiplets and can be as low as TeV if a small Yukawa coupling $\kappa$ is allowed.

The new fermionic multiplets in TeV Seesaw mechanisms can be experimentally accessible at the Large Hadron Collider (LHC) (see Ref.~\cite{Cai:2017mow} and the references therein). Up to now, no significant evidence of such heavy fermions was observed at the LHC and the lower limits on their masses are highly dependent on the mixing between the heavy Majorana neutrinos and the SM leptons.
A majority of search results released by ATLAS~\cite{Aad:2015xaa,Aad:2015cxa} and CMS~\cite{Khachatryan:2015gha,Khachatryan:2016olu,Sirunyan:2018mtv} were defined by the analysis of electron and/or muon final states from heavy neutrino decay~\footnote{There were searches for heavy neutrinos using events with two tau leptons and two jets in a left-right symmetric model~\cite{Sirunyan:2017yrk,Sirunyan:2018vhk}.}. However, governed by the constraints from neutrino oscillation experiments, the channel with tau lepton in heavy neutrino decay also plays an important role in confirming neutrino mass hierarchies and determining mixing parameters~\cite{Perez:2008zc,Perez:2008ha,Li:2009mw,Cai:2017mow,Li:2018jns,Pascoli:2018rsg,CidVidal:2018eel,Pascoli:2018heg,Cheung:2020buy,Ashanujjaman:2021jhi,Das:2020uer}.
Recently, several precise reactor measurements lead to more accurate lepton flavor predictions about the Seesaw models. For instance, Double Chooz~\cite{Abe:2011fz}, RENO~\cite{Ahn:2012nd} and in particular Daya Bay~\cite{An:2012eh}, have reported non-zero measurements of $\theta_{13}$ by looking for the disappearance of anti-electron neutrino. T2K and NOvA reported on indications of a non-zero leptonic CP phase~\cite{Abe:2013hdq,Adamson:2016tbq,Abe:2017uxa}. These experiments provide us up-to-date neutrino oscillation results
to investigate the impact on neutrino mass models and consequently examine the lepton flavor signatures to be searched at colliders.
Thus, from the experimental point of view, a detailed assessment of the search sensitivity to the channels with tau lepton is in demand at the LHC upgrades.

On the other hand, as proved in Refs.~\cite{Kersten:2007vk,Moffat:2017feq}, requiring all three light neutrinos to be massless is equivalent to requiring lepton number to be conserved at all orders in perturbation theory. In other words, lepton number is nearly conserved in low-scale seesaw models with fermionic singlets in light of non-zero neutrino masses. Tiny neutrino mass is proportional to small lepton number violation (LNV) parameters. Thus, for low-scale Type I Seesaw, the observable lepton number violation is decoupled in high energy processes. Any observation of lepton number violation at colliders implies an extension of the canonical Type I Seesaw. Below we consider the introduction of a new U$(1)_{\rm B-L}$ gauge symmetry in which the SM matter fields and heavy neutrinos are charged under the B-L (baryon minus lepton number) symmetry. However, the minimal B-L model is highly constrained by the di-lepton search at the LHC. The lower $Z'$ mass is pushed up to 4.2 TeV for the minimal U$(1)_{\rm B-L}$ model~\cite{Aaboud:2017buh}. To evade the di-lepton constraint, the minimal B-L model should be extended. In a flavored U$(1)_{\rm B-L}$ model (denoted by U$(1)_{\rm (B-L)_3}$) where only the third generation fermions couple to the new gauge boson $Z'$~\cite{Alonso:2017uky,Cox:2017eme}, the constraint on the new gauge interaction is much weak and a single heavy neutrino preferably decays into the tau lepton. In addition, leptogenesis prefers much heavy right-handed neutrinos and the usual B-L breaking scale should be high. However, it is sufficient to have two heavy right-handed neutrinos to explain the baryon asymmetry of the Universe, which leaves a possibility that the third right-handed neutrino could be light enough to be probed at future colliders. It is thus plausible to consider this non-universal U$(1)_{\rm (B-L)_3}$ model.
We study the pair production of heavy neutrino via gauge interaction in the U$(1)_{\rm (B-L)_3}$ extended Type I Seesaw~\cite{Perez:2009mu,Cox:2017eme,Chiang:2019ajm}
\begin{eqnarray}
pp\to Z'\to NN\;,
\end{eqnarray}
where $Z'$ is the new gauge boson of the U$(1)_{\rm (B-L)_3}$ symmetry and $N$ denotes the heavy neutrino followed by the decay into tau lepton in LNV final states.
Suppose a new $Z'$ can be discovered through the di-lepton channel in future, one can look for the LNV signal and test the specific nature of heavy neutrinos through this channel.
We will show the sensitivity of the search for the above LNV channels in both the high-luminosity LHC (HL-LHC) and future circular hadron collider (FCC-hh).

This paper is organized as follows.
In Sec.~\ref{sec:Seesaw}, we first describe the U$(1)_{\rm B-L}$ extension and discuss the realization of Type I Seesaw in this model and the constraint from neutrino oscillation experiments on heavy neutrino decay patterns. In Sec.~\ref{sec:Search}, we simulate the pair production of heavy neutrinos via $Z'$ and its LNV signature with tau lepton at the LHC upgrade.
The results of projected sensitivity for heavy neutrinos using tau lepton are also given.
Finally, in Sec.~\ref{sec:Concl} we summarize our conclusions.

\section{Type I Seesaw in U$(1)_{\rm B-L}$ extension model}
\label{sec:Seesaw}

\subsection{U$(1)_{\rm B-L}$ extension model}
To reveal the origins of the B-L accidental symmetry in the SM and Majorana neutrino masses, one considers the extension of the SM with U$(1)'$ as a local symmetry.
In the well-studied U$(1)'$ Abelian gauge extension of the SM~\cite{Appelquist:2002mw}, U$(1)'$ is a linear combination of U$(1)_{\rm Y}$ and U$(1)_{\rm B-L}$ after the spontaneous breaking of electroweak symmetry and B-L symmetry.
In this model, right-handed neutrinos (RH) are introduced to cancel gauge anomalies and realize the Type I Seesaw mechanism. The relevant covariant derivatives can be generally written as
\begin{eqnarray}
D_\mu \ni ig_1 Y B_\mu + i(g_1' Y+g' Y_{BL})B'_\mu = ig_1 Y B_\mu + i g_X (x_1' Y+x' Y_{BL})B'_\mu,
\end{eqnarray}
where $B_\mu (Y)$ and $B'_\mu (Y_{BL})$ are the gauge fields (quantum numbers) of U$(1)_{\rm Y}$ and U$(1)_{\rm B-L}$, respectively.
Here we redefine the gauge coupling of $B'_\mu$ as $g_X\equiv g_1'/x_1'=g'/x'$ with U$(1)'$ charge $x_1' Y+x' Y_{BL}$. Besides the gauge field $B'_\mu$, a new scalar field $S\sim (1,1,0,2x')$ is introduced to break the local U$(1)_{\rm B-L}$ symmetry and generate the masses of right-handed neutrinos $N_R\sim (1,1,0,-x')$. When $x'=1, x_1'=0$, the mixing between U$(1)_{\rm Y}$ and U$(1)_{\rm B-L}$ vanishes and the most economical extension emerges~\cite{Davidson:1978pm,Mohapatra:1980qe,Marshak:1979fm,Wetterich:1981bx,Masiero:1982fi,Mohapatra:1982xz,Buchmuller:1991ce}, i.e. U$(1)'=$U$(1)_{\rm B-L}$ and $g_X=g'$. In the following studies, we consider this economical U$(1)_{\rm B-L}$ extension.
Once $S$ gets the vacuum expectation value (vev) $\langle S\rangle=v_S/\sqrt{2}$, the B-L symmetry is broken and the new gauge boson $Z'=Z_{\rm X}$ can also get mass $M_{Z'}=2g' v_S$ from the kinetic term $(D_\mu S)^\dagger (D^\mu S)$ with $D_\mu S=\partial_\mu S+i2g'B'_\mu S$. LEP-II sets the lower bound on the $Z'$ boson mass in U$(1)_{\rm B-L}$ gauge model, i.e. $M_{Z'}/g'\gtrsim 6$ TeV~\cite{Carena:2004xs}. The most stringent limit comes from the search for dilepton resonance at the LHC with $\sqrt{s}=13$ TeV and the luminosity of 36 fb$^{-1}$~\cite{Aaboud:2017buh} or 139 fb$^{-1}$~\cite{Aad:2019fac}. Whereas, lower mass limit is set up to 4.2 TeV for the minimal U$(1)_{\rm B-L}$ model~\cite{Aaboud:2017buh} and the constraint is also given in the plane of coupling strength $g'/g_Z$ vs. $M_{Z'}$.

On the other hand, in this model, the anomaly-free U$(1)'$ extension allows non-universal U$(1)'$ charges for the SM fermions.
With the presence of three right-handed neutrinos, the most general requirement of the U$(1)'$ charges is~\cite{Kownacki:2016pmx}
\begin{eqnarray}
3(Q'_1+Q'_2+Q'_3)+Q'_e+Q'_\mu+Q'_\tau=0\;,
\end{eqnarray}
where $Q'_{1,2,3}$ are the quark charges and $Q'_{e,\mu,\tau}$ are the lepton charges.
To relax the constraint from dilepton channel, there proposed a flavored model under the U$(1)_{\rm (B-L)_3}$ symmetry where only the third generation fermions are charged with $3 Q'_3=-Q'_\tau=1$ and $Q'_1=Q'_2=Q'_e=Q'_\mu=0$~\cite{Alonso:2017uky,Cox:2017eme}. The $Z'$ can only be produced through $b\bar{b}\to Z'$ and decay into $\tau^+ \tau^-$ and a pair of one single heavy neutrino. As a result, the above dilepton constraint can be evaded. The constraint from di-tau search~\cite{Aaboud:2017sjh} is expected to be rather weak~\cite{Han:2019zkz}.
Note that the constraint can also be relaxed by maximizing the ratio $\Gamma(Z'_{BL}\to \ell^+\ell^-)/\Gamma(Z'_X\to \ell^+\ell^-)$~\cite{Das:2017flq,Das:2017deo,Das:2019fee} with $Z'_{BL}$ and $Z'_X$ being the $Z'$ gauge bosons under the minimal U$(1)_{\rm B-L}$ model and the general U$(1)'$ model, respectively. Below we take the U$(1)_{\rm (B-L)_3}$ model for illustration although our result below is generic and applicable to any U$(1)'$ model with heavy neutrinos.

\subsection{Type I Seesaw and heavy neutrino decay}

The neutrino Yukawa interactions are
\begin{eqnarray}
-\mathcal{L}_Y^I=Y^D_\nu\bar{l}_L\tilde{H}N_R+{1\over 2}Y^M_\nu\overline{(N^c)_L} N_R S +h.c.,
\end{eqnarray}
where $\tilde{H}=i\sigma_2 H^\ast$ with $H$ being the SM Higgs doublet. The right-handed neutrino mass matrix is spontaneously generated to be $M_R=Y^M_\nu v_S/\sqrt{2}$. The interaction for $Z'$ and heavy neutrinos is obtained from the kinetic term $i\overline{N_R}\gamma^\mu D_\mu N_R$ with $D_\mu N_R=\partial_\mu N_R-ig'B'_\mu N_R$.
With the right-handed neutrino mass matrix $M_R$ obtained above, the Seesaw formula and the mixing between the SM charged leptons and heavy neutrinos
in this model follow the same as those in the canonical Type I Seesaw. After $H$ receives the vev $\langle H\rangle=v/\sqrt{2}$
and the Dirac masses $m_D=Y^D_\nu v/\sqrt{2}$ appear, one has the diagonalization of the neutrino mass matrix
\begin{eqnarray}
\mathbb{N}^\dagger
\left(
  \begin{array}{cc}
    0 & m_D \\
    m_{D}^{T} & M_R \\
  \end{array}
\right) \mathbb{N}^\ast &=&
\left(
  \begin{array}{cc}
    m_\nu & 0 \\
    0 & M_N  \\
  \end{array}
\right),
\label{eq:type1NuMixMatrix}
\end{eqnarray}
and the eigenstate transformation
\begin{eqnarray}
\left(
  \begin{array}{c}
    \nu_L \\
    (N^c)_L \\
  \end{array}
\right) = \mathbb{N}\left(
  \begin{array}{c}
    \nu_L \\
    (N^c)_L \\
  \end{array}
\right)_{mass}, \ \ \ \mathbb{N}= \left(
  \begin{array}{cc}
    U & V \\
    X & Y \\
  \end{array}
\right)\;.
\label{eq:nuMixDefs}
\end{eqnarray}
With a matrix $E$ diagonalizing the charged lepton mass matrix, we can define
\begin{eqnarray}
E^\dagger U\equiv U_{PMNS}\;,~~~E^\dagger V\equiv V_{\ell N}\;,
\end{eqnarray}
where $U_{PMNS}$ is the approximate Pontecorvo-Maki-Nakagawa-Sakata (PMNS) neutrino mass mixing matrix and the matrix $V_{\ell N}$
transits heavy neutrinos to charged leptons~\cite{Atre:2009rg}.

One can derive a relation between diagonalized neutrino mass matrices and mixing matrices
\begin{eqnarray}
U^\ast_{PMNS}m_\nu^{diag} U^\dagger_{PMNS}+V^\ast_{\ell
N}M_N^{diag}V^\dagger_{\ell N}=0 \ ,
\label{typei}
\end{eqnarray}
with mass eigenvalues $m_\nu^{diag}=(m_1,m_2,m_3)$ and $M_N^{diag}=(M_{N_1},M_{N_2},M_{N_3})$.
Here the masses and mixing of the light neutrinos in the first term are measurable from the neutrino oscillation experiments,
and the second term contains the masses and mixing of the new heavy neutrinos.
The general solution of the $V_{\ell N}$ in Eq.~(\ref{typei}) can be parameterized in terms of an arbitrary orthogonal complex matrix $\Omega$, i.e. the Casas-Ibarra parametrization~\cite{Casas:2001sr}
\begin{eqnarray}
V_{\ell N}=U_{PMNS} (m_\nu^{diag})^{1/2}\Omega (M_N^{diag})^{-1/2},
\label{omega}
\end{eqnarray}
with the orthogonality condition $\Omega \Omega^T=I$.
Using the SM electroweak current for heavy Majorana neutrinos $N_i$, in the mixed mass-flavor basis, we can obtain the partial width of their decay into charged lepton
\begin{eqnarray}
\Gamma(N_i\to \ell^\pm W^\mp)={G_F\over 8\sqrt{2}\pi}|V_{\ell N_i}|^2 M_{N_i}(M_{N_i}^2+2M_W^2)\left(1-{M_W^2\over M_{N_i}^2}\right)^2,
\label{partialN}
\end{eqnarray}
and the asymptotic behavior holds when $M_{N_i}\gg M_W, M_Z, M_h$.
\begin{eqnarray}
\Gamma(N_i\to \sum_\ell \ell^\pm W^\mp)\approx \Gamma(N_i\to \sum_\nu \nu Z+\bar{\nu}Z)\approx \Gamma(N_i\to \sum_\nu \nu h+\bar{\nu}h)\;,
\end{eqnarray}
where $\ell=e,\mu,\tau$.
Note that, more generally, the Dirac mass matrix can be quite arbitrary with three complex angles parameterizing the orthogonal matrix $\Omega$~\cite{Perez:2009mu,FileviezPerez:2020cgn}. The above asymptotic behavior of heavy neutrino decay branching ratios is consistent with the Goldstone equivalence theorem and indicates BR$(N_i\to \sum_\ell \ell^\pm W^\mp)$ is nearly equal to $25\%$ for any heavy neutrino $N_i$. The branching fraction of $N_i~(i=1,2,3)$ decays into each lepton flavor is thus less than $25\%$ for any possible $\Omega$ matrix. In one simple case for $\Omega$, i.e. a diagonal unity matrix $\Omega=I$, the results of $|V_{\ell N}|^2$ are proportional to one and only one light neutrino mass~\cite{Perez:2009mu}. The branching ratio of $N_{i}\to \ell^\pm W^\mp$ for each lepton flavor is thus independent of neutrino mass~\cite{Perez:2009mu}. The representative choice of $\Omega=I$ can also lead to both the leading and sub-leading decay rate cases in each heavy neutrino decay into certain charged lepton flavor, so one can distinguish different heavy neutrinos~\cite{FileviezPerez:2020cgn}.
Thus, to reduce the parameter dependence and obtain certain predictions, we take this illustrative case of Casas-Ibarra parametrization with $\Omega=I$ in the following study.

In order to understand the implication of the neutrino experiments,
we then discuss the neutrino mass and mixing parameters in light of oscillation data.
The neutrino mixing matrix can be parameterized as
\beq U_{PMNS}= \left(
\begin{array}{lll}
 c_{12} c_{13} & c_{13} s_{12} & e^{-\text{i$\delta $}} s_{13}
   \\
 -c_{12} s_{13} s_{23} e^{\text{i$\delta $}}-c_{23} s_{12} &
   c_{12} c_{23}-e^{\text{i$\delta $}} s_{12} s_{13} s_{23} &
   c_{13} s_{23} \\
 s_{12} s_{23}-e^{\text{i$\delta $}} c_{12} c_{23} s_{13} &
   -c_{23} s_{12} s_{13} e^{\text{i$\delta $}}-c_{12} s_{23} &
   c_{13} c_{23}
\label{PMNS}
\end{array}
\right)\times \text{diag} (e^{i \Phi_1/2}, 1, e^{i \Phi_2/2}) \eeq
where $s_{ij}\equiv\sin{\theta_{ij}}$, $c_{ij}\equiv\cos{\theta_{ij}}$, $0 \le
\theta_{ij} \le \pi/2$ and $0 \le \delta, \Phi_i \le 2\pi$ with $\delta$ being the Dirac CP phase and $\Phi_i$ the Majorana phases.
The size of the mass-squared splitting between three
neutrino states is extracted from neutrino oscillation experiments.
The sign of $\Delta m_{31}^2 = m_{3}^2 - m_{1}^2$, however, still remains unknown, which can be either positive, the
Normal Hierarchy (NH), or negative, the Inverted Hierarchy (IH), for the spectrum of the neutrino masses.
Taking into account the data on atmospheric neutrinos provided by the Super-Kamiokande collaboration,
the latest best global fit results of the neutrino masses and mixing parameters are as below~\cite{Esteban:2020cvm,nufit2020}
\begin{eqnarray}
\Delta m_{21}^2 &=&  7.42 \times 10^{-5} \ev^2 \;, \nonumber \\
\Delta m_{31}^2 &=&  2.517 \times 10^{-3} \ev^2 \ (\Delta m_{32}^2  =  -2.498 \times 10^{-3} \ev^2 )\;, \nonumber \\
\sin^2{\theta_{12}} &=&  0.304\;,~~\sin^2{\theta_{23}} =  0.573 \ (0.575)\;, \nonumber \\
\sin^2{\theta_{13}} &=& 0.02219 \ (0.02238) \;,~~\delta_{\rm CP} = 197^\circ (282^\circ)\;,
\end{eqnarray}
for NH (IH).
In addition, the sum of neutrino masses is constrained
by combining the Planck, WMAP, highL and BAO data~\cite{Ade:2015xua} at $95\%$ confidence level (CL) as,
\beq
\sum_{i=1}^3 m_i < \ 0.230 \ \ev .
\eeq
By inputting the above experimental constraints to the transiting matrix $V_{\ell N}$ and the above decay width formulas, we can obtain the preferred values for
heavy neutrino decay patterns below.
In our numerical calculations below, we take the benchmark decay branching ratios of heavy neutrinos with $\Omega=I$ for NH and IH as shown in Table~\ref{BR}.

\begin{table}[tb]
\begin{center}
\begin{tabular}{|c|c|c|c|}
\hline
BR$(N_i)$  & $e^\pm W^\mp$ & $\mu^\pm W^\mp$ & $\tau^\pm W^\mp$
\\ \hline
$N_1$ & 17\% (17\%) & 1.85\% (3.8\%) & 6.15\% (4.2\%)
\\ \hline
$N_2$ & 7.43\% (7.43\%) & 9.15\% (7.14\%)  & 8.42\% (10.43\%)
\\ \hline
$N_3$ & 0.55\% (0.56\%) & 14\% (14.04\%)  & 10.45\% (10.4\%)
\\ \hline
\end{tabular}
\end{center}
\caption{Benchmark decay branching ratios of $N_i$ in $\Omega=I$ case for NH (IH). Here we assume the lowest light neutrino mass as $10^{-4}$ eV and $M_{N_i}\gg M_W$.}
\label{BR}
\end{table}

\begin{figure}[ht]
\centering
\includegraphics[width=0.48\textwidth]{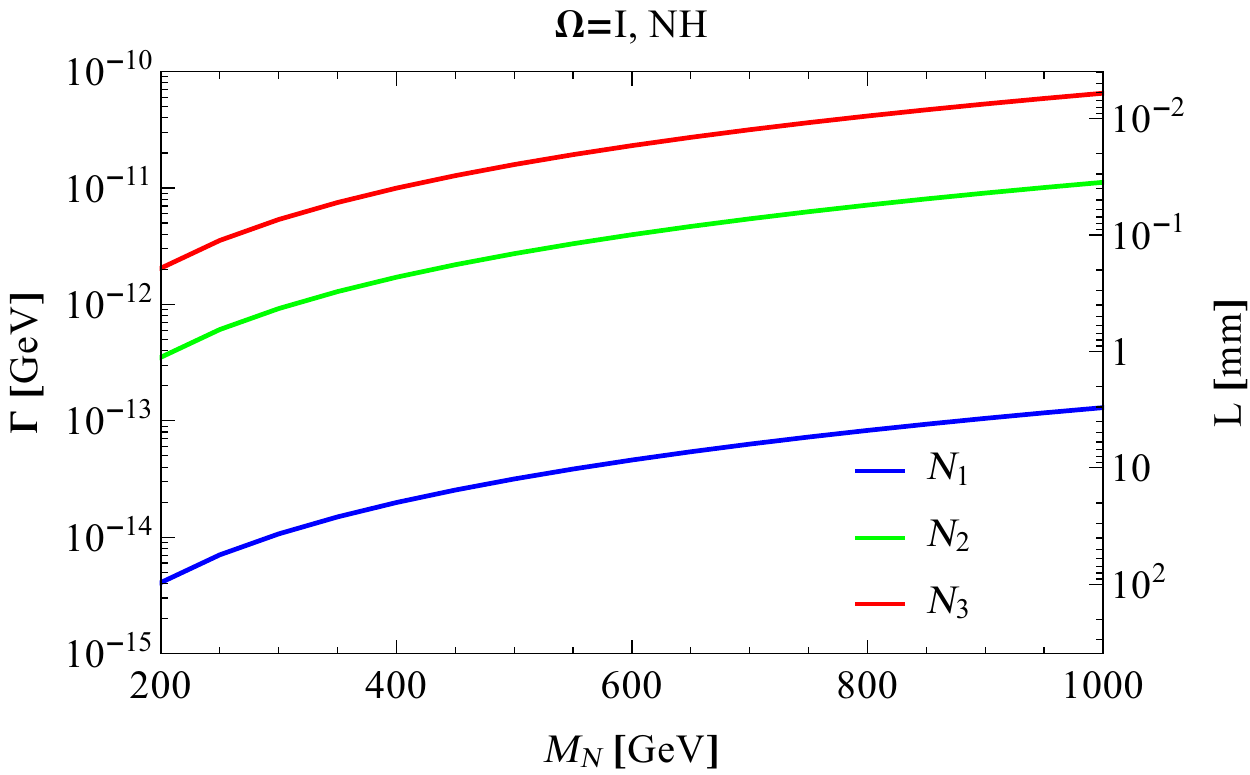}
\includegraphics[width=0.48\textwidth]{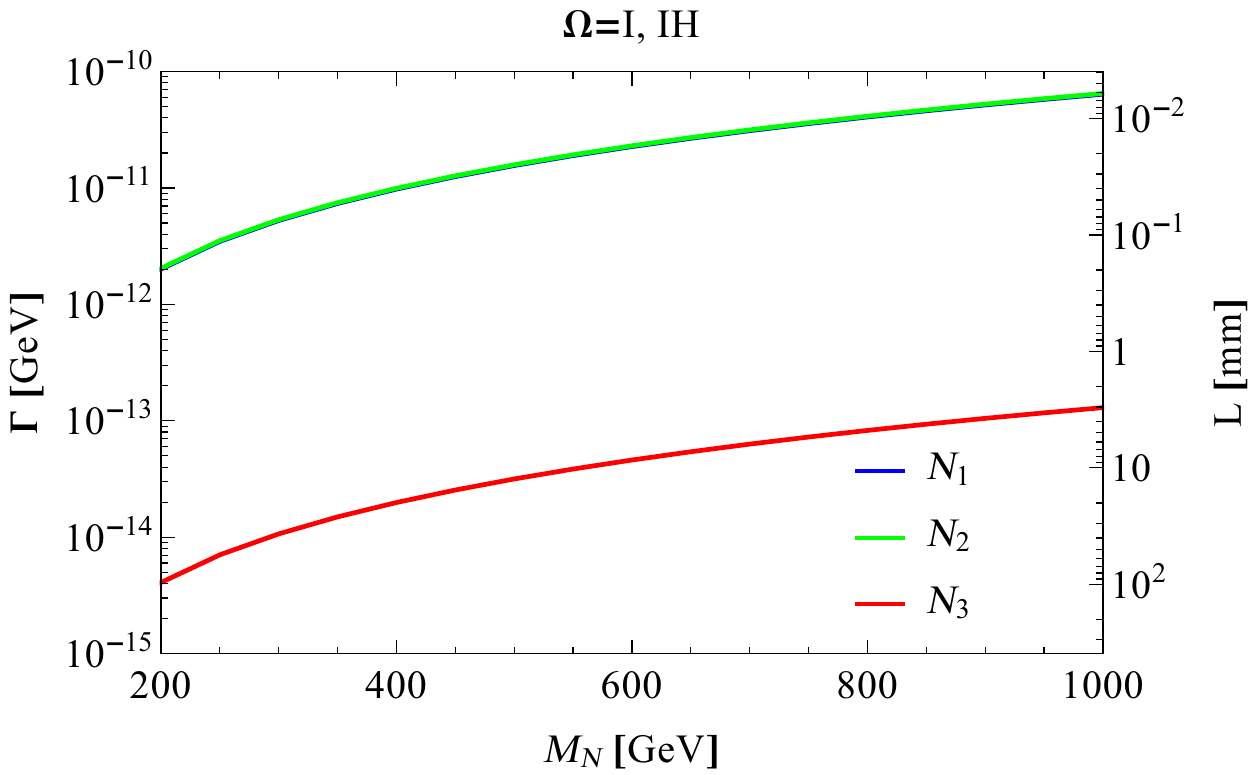}
\caption{The total width (left axis) and decay length (right axis) versus $M_N$ for $N_i (i=1,2,3)$ and NH (left) and IH (right) scenarios. We assume $M_N=M_{Z'}/4$ and $\Omega=I$.}
\label{fig:lifetime}
\end{figure}

The total decay widths of heavy Majorana neutrinos are proportional to $|V_{\ell N}|^2\sim m_\nu/M_N$ which is small and may lead to long decay length.
We calculate the decay length of the Majorna neutrinos by the formula $L= \frac{1}{\Gamma} \frac{E_N}{M_N}$ where $\frac{E_N}{M_N}$ is the boost factor and $E_N$ is roughly considered to be $M_{Z^\prime}/2$ in the rest frame of $Z'$. Fig.~\ref{fig:lifetime} shows the total width (left axis) and decay length (right axis) versus $M_N$ for $N_i~(i=1,2,3)$ and $\Omega=I$. We find that in most of cases the decay length is less than $\mathcal{O}(1)$ mm and the heavy neutrinos can be viewed as decaying promptly. Only for the cases of $N_1$ in NH and $N_3$ in IH, the typical decay length is above 2 mm. In particular, for 250~GeV$< M_N < 400$~GeV, the decay length could vary from 10~mm to 30~mm. The decay length is sufficiently long to perform the displaced vertex searches. This is a clear difference between the NH and IH scenarios as well as different heavy neutrinos. If this is observed in future, it could serve as an indication to distinguish the neutrino patterns. Despite of this, such length is still short compared with the size of inner detector of ATLAS (1.2 meter in radius) and the prompt decay approximation is still valid. We leave such interesting possibilities for future studies and in the following analysis we assume the Majorana neutrinos promptly decay.

\section{Searching for heavy neutrinos with tau lepton at hadron colliders}
\label{sec:Search}

\begin{figure}[ht]
\centering
\includegraphics[width=0.4\textwidth]{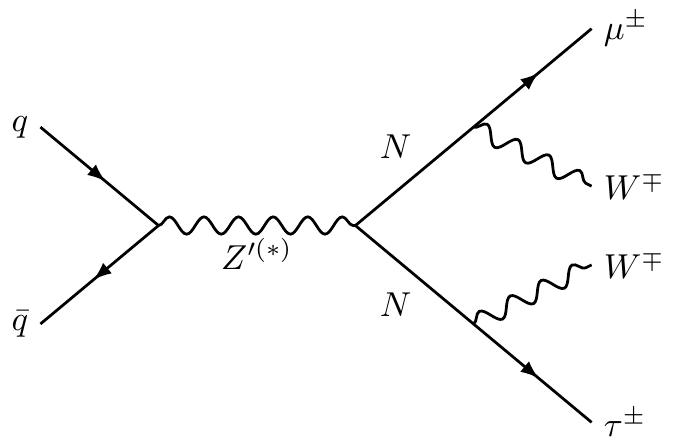}
\caption{The pair production of heavy neutrino $N$ through $Z'$ resonance. The fermion arrows in final states are only for $\mu^-\tau^-$ and the charge conjugation should be also accounted.}
\label{fig:diagram_typeI}
\end{figure}

The interesting production at hadron colliders for
heavy neutrinos in U$(1)'$ extended Type I Seesaw is the pair production of heavy neutrinos through $Z'$ resonance as shown in Fig.~\ref{fig:diagram_typeI}.
In the U$(1)_{\rm (B-L)_3}$ model, the most appealing channel is
\begin{eqnarray}
b\bar{b}\to Z' \to NN \to \ell^\pm \tau^\pm W^\mp W^\mp\;,
\end{eqnarray}
where $\ell=e, \mu$ and the heavy neutrino $N$ can be any one of the mass eigenstates in Table~\ref{BR} as they can all involve the third generation coupled to $Z'$.
The signal channel we consider includes one electron or muon and one same-sign hadronically decaying tau lepton in the subsequent decay of $N$. The same-sign $W$ bosons are required to decay hadronically. The U$(1)_{\rm (B-L)_3}$ model file is produced by FeynRules~\cite{Alloul:2013bka} and is interfaced with MadGraph5\_aMC@NLO~\cite{Alwall:2014hca} to generate signal events. The major irreducible SM background is from
$t\bar{t}W^\pm\to b\bar{b}W^\pm W^\pm W^\mp$ with the two same-sign $W$ bosons decaying to one charged lepton $\ell=e, \mu$ and one tau lepton, the opposite-sign $W$ hadronically decays. The SM backgrounds also include $t\bar{t}Z\to b\bar{b}W^+ W^- Z$ and $WZ$+jets with the $Z$ boson leptonic decay.
Both the signal and background events are then passed to Pythia 8~\cite{Sjostrand:2014zea} and Delphes 3~\cite{deFavereau:2013fsa} for parton shower and detector simulation, respectively.

\begin{figure}[htb!]
\centering
\includegraphics[width=0.47\textwidth]{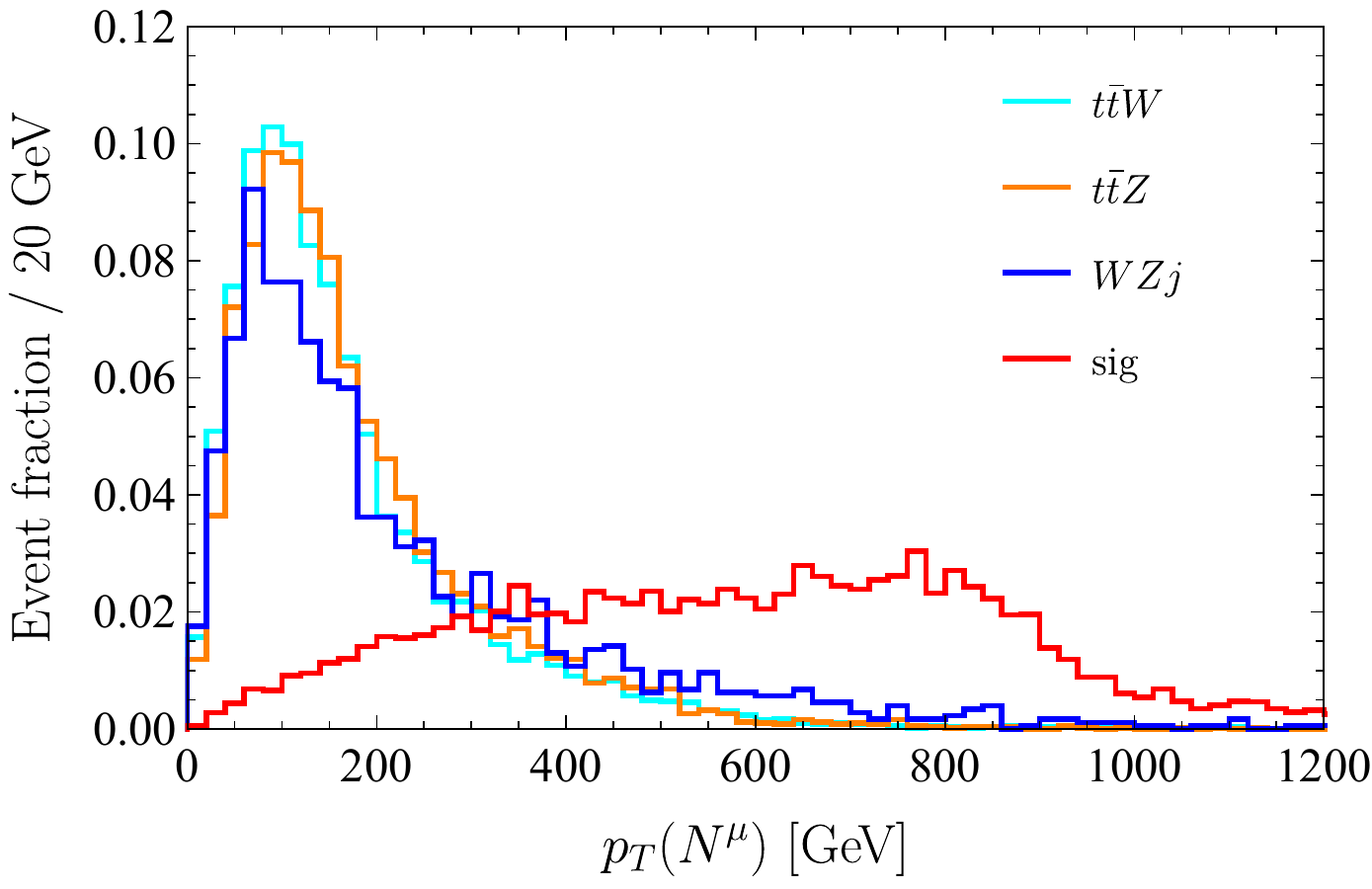}
\includegraphics[width=0.47\textwidth]{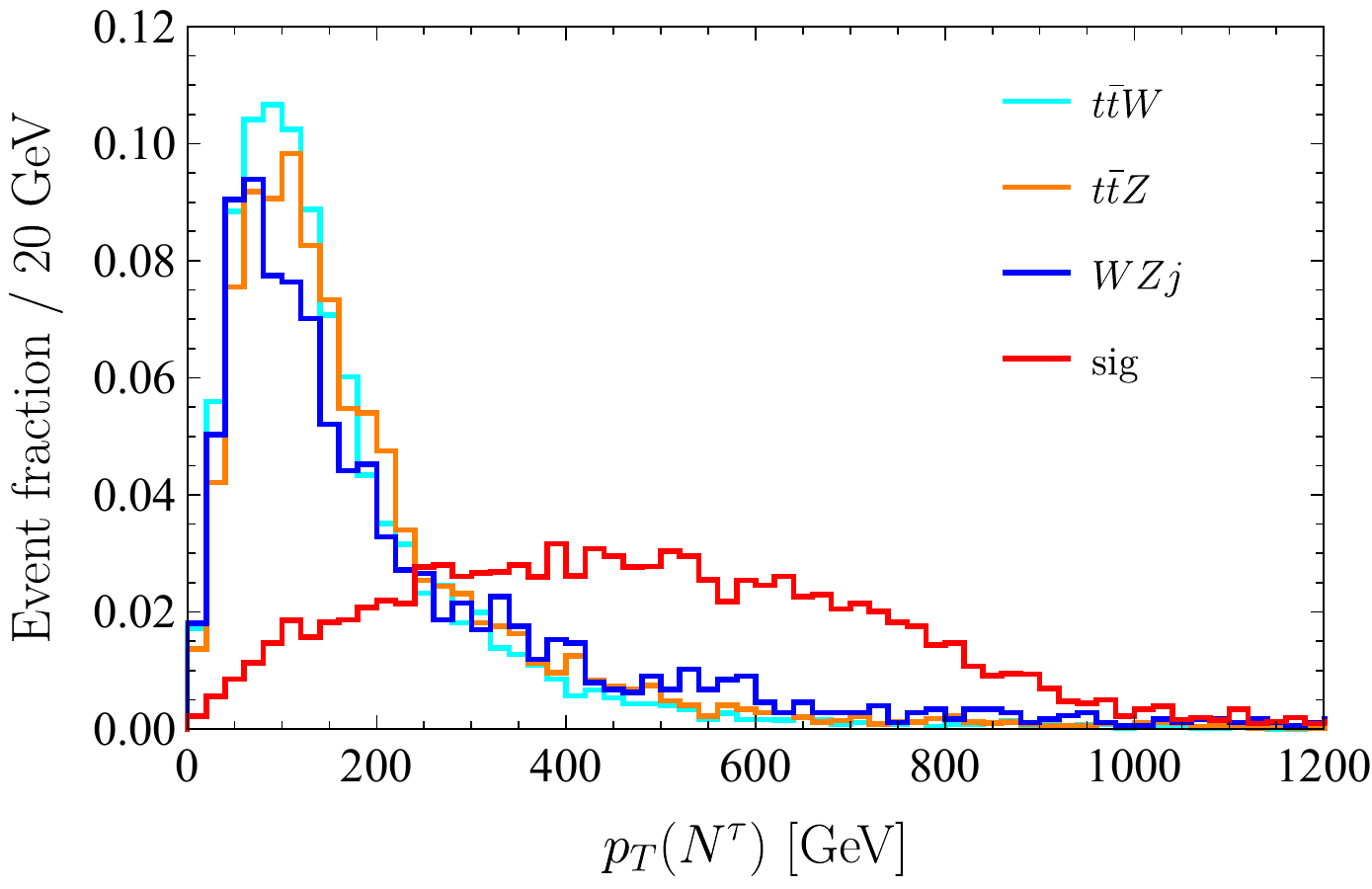}\\
\includegraphics[width=0.46\textwidth]{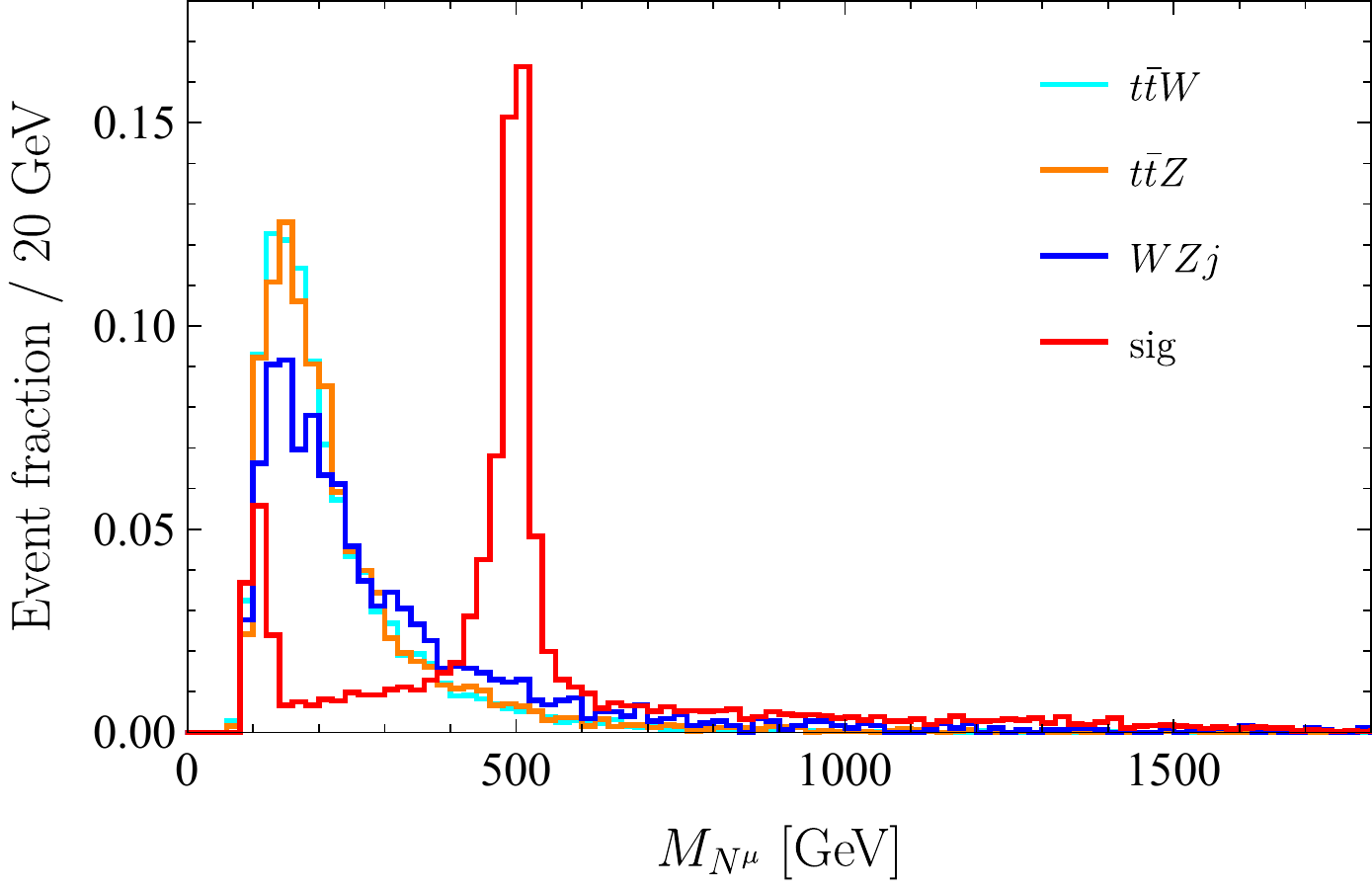}
\includegraphics[width=0.46\textwidth]{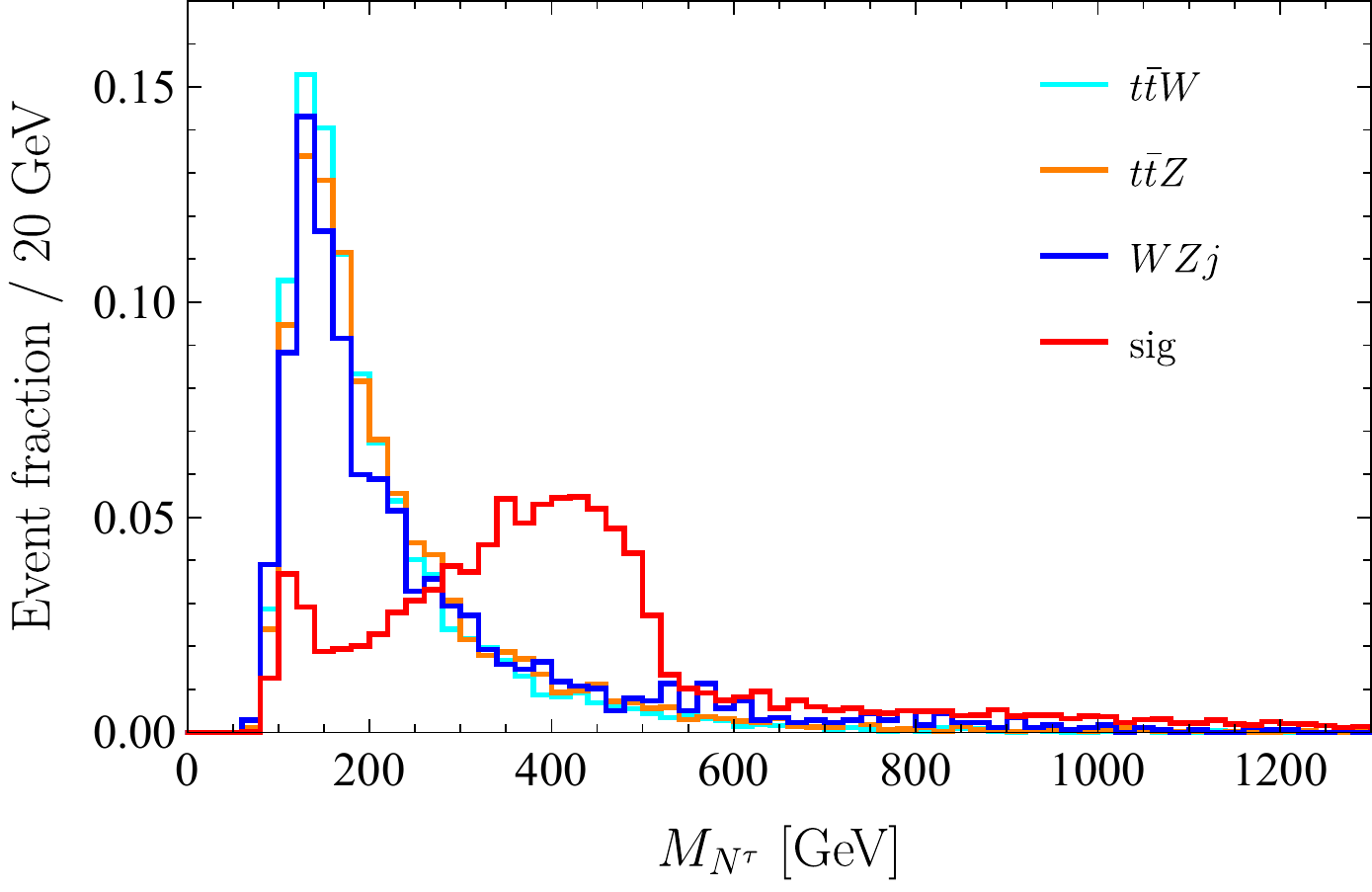}
\caption{Normalized distributions of $p_T$ (top) and invariant mass (bottom) for reconstructed $N^\mu$ (left) and $N^\tau$ (right). We assume $M_{Z^{\prime}}=2$~TeV and $M_{N}=M_{Z^{\prime}}/4$ for illustration. }
\label{fig:pt_NR}
\end{figure}

We select the events with exactly one electron or muon and one same-sign tau lepton satisfying
\begin{eqnarray}
|\eta(\ell)|< 2.4,~p_T(\ell)>25~{\rm GeV};~~~|\eta(\tau)|< 2.5,~p_T(\tau)>25~{\rm GeV}\;.
\end{eqnarray}
The jets from $W$ decay are clustered using the anti-$k_T$ algorithm in FastJet~\cite{Cacciari:2011ma}. If the jets are well isolated, we reconstruct each $W$ boson from two resolved jets with $|\eta(j)|< 2.4$ and $p_T(j)>25$~GeV. As we consider the decay of heavy neutrino, the $W$ boson can be boosted and the decay products are likely to merge into a single fat jet $J$~\cite{Das:2020gnt}. The fat jet is reconstructed via the anti-$k_T$ algorithm with $R = 0.8$. The fat jet with $65<M_J<95$~GeV, $p_T(J)>250$~GeV and $|\eta(J)|<2.5$ is identified as boosted $W$ candidate. If there are more than two boosted $W$ candidates, we sort them by $|M_J-M_{W}|$ and select the first two as those from $N$ decay.
If there is only one boosted $W$ boson, the resolved jets are required to satisfy $\Delta R(W, j)>0.8$ and the other $W$ boson
candidate is reconstructed by choosing the pair of jets which
minimizes $|M_{jj}-M_W|$. The di-jet invariant mass should satisfy
$50<M_{jj}<110$~GeV in order to be identified as a $W$ boson.
If there is no boosted $W$ boson, there should be at least two
pairs of resolved jets to reconstruct the two $W$ bosons. We sort their combinations by $|M_{jj}-M_W|$
and select the first two to construct $W$ bosons.
Given the same-sign $\ell$ and $\tau$ and the two reconstructed $W$ bosons, we find the combination that minimizes
$\Delta R(\ell,W_1)+\Delta R(\tau,W_2)$ and then construct $N^{\ell}$ and $N^{\tau}$ respectively denoting the heavy neutrino decay into $\ell$ and $\tau$. In Fig.~\ref{fig:pt_NR}, assuming $M_{Z^{\prime}}=2$~TeV and $M_{N}=M_{Z^{\prime}}/4$, the distributions of $p_T$ and invariant mass are respectively shown for the two reconstructed $N$.
We can find that $M_{N^\mu}$ peaks at 500 GeV which is very
close to the true value, while the reconstruction of $N$ from $(\tau,W)$ is not as good as that from
$(\mu,W)$. As a result, in the final selection, we only
apply the minimal $p_T$ cut and the mass window for $N^{\mu}$
\begin{eqnarray}
p_T(N^{\mu})>250~{\rm GeV},~~p_T(N^{\tau})>200~{\rm GeV};~~~|M_{N^{\mu}}-M_{N}|<0.1 M_{N}\;.
\end{eqnarray}
The cuts for $M_{N^{\tau}}$ and the reconstructed $Z^{\prime}$ are not applied.

\begin{figure}[htb!]
\centering
\minigraph{7cm}{-0.05in}{(a)}{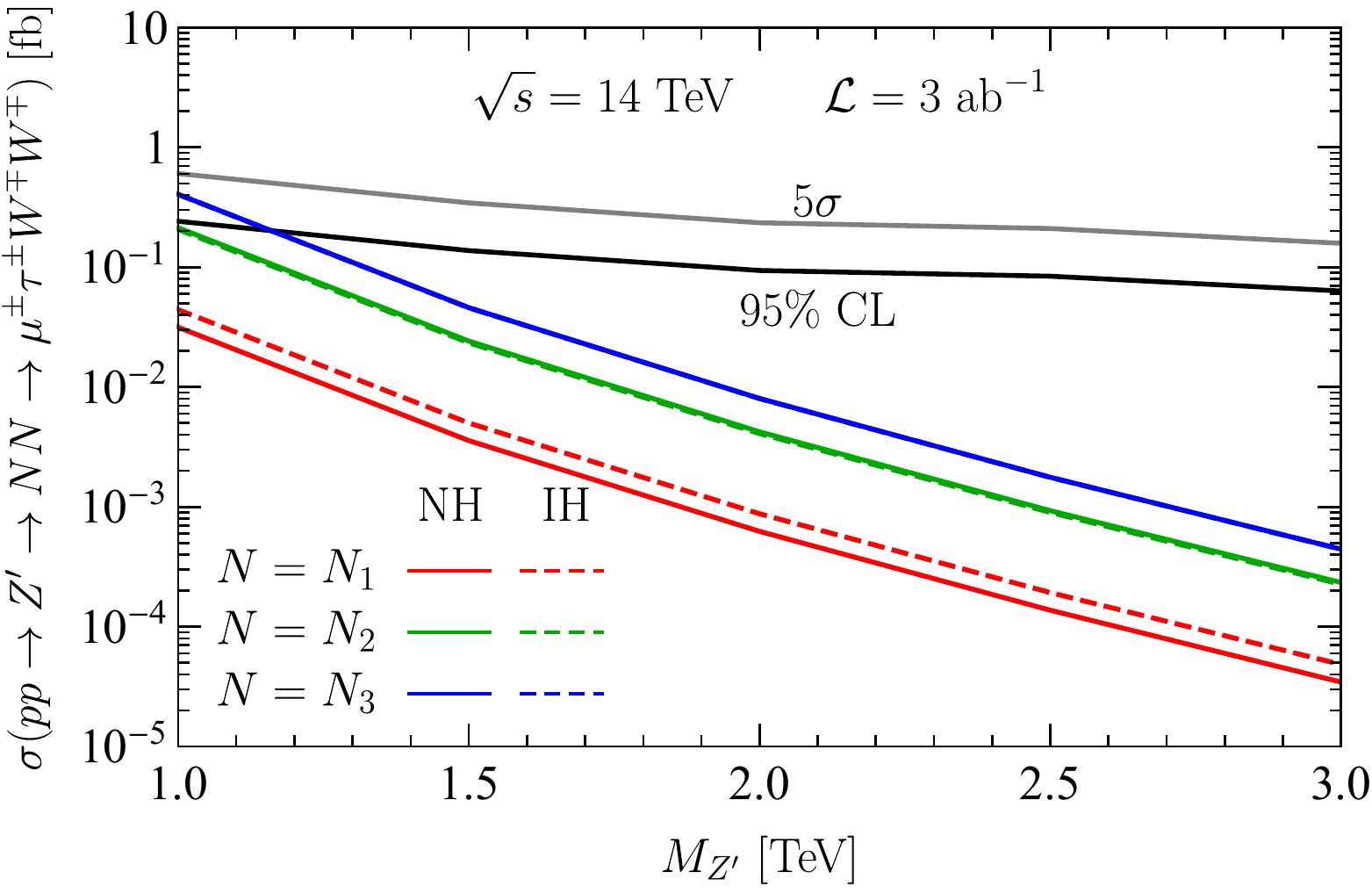}\\[0.2cm]
\minigraph{7cm}{-0.05in}{(b)}{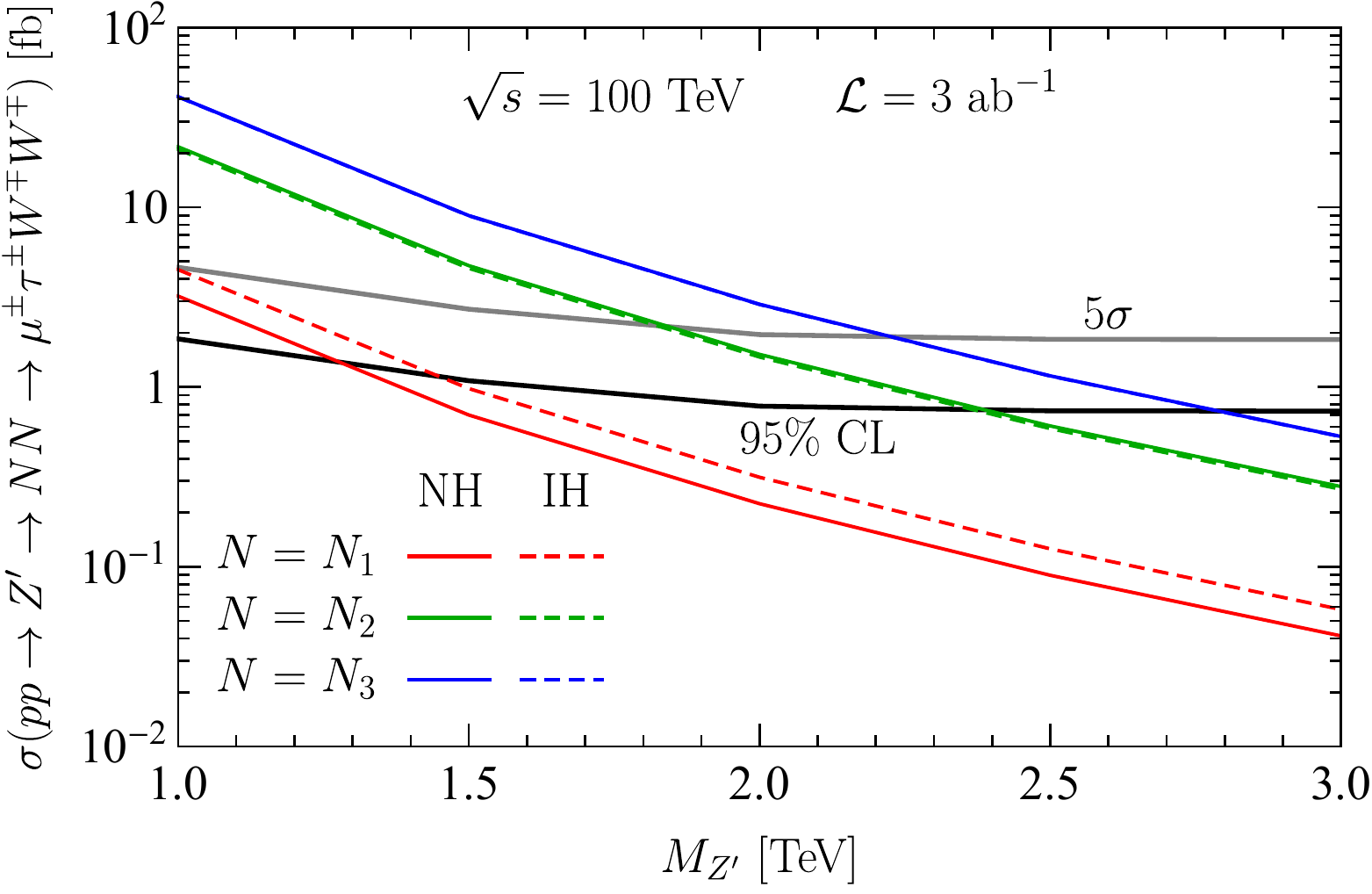}
\minigraph{7cm}{-0.05in}{(c)}{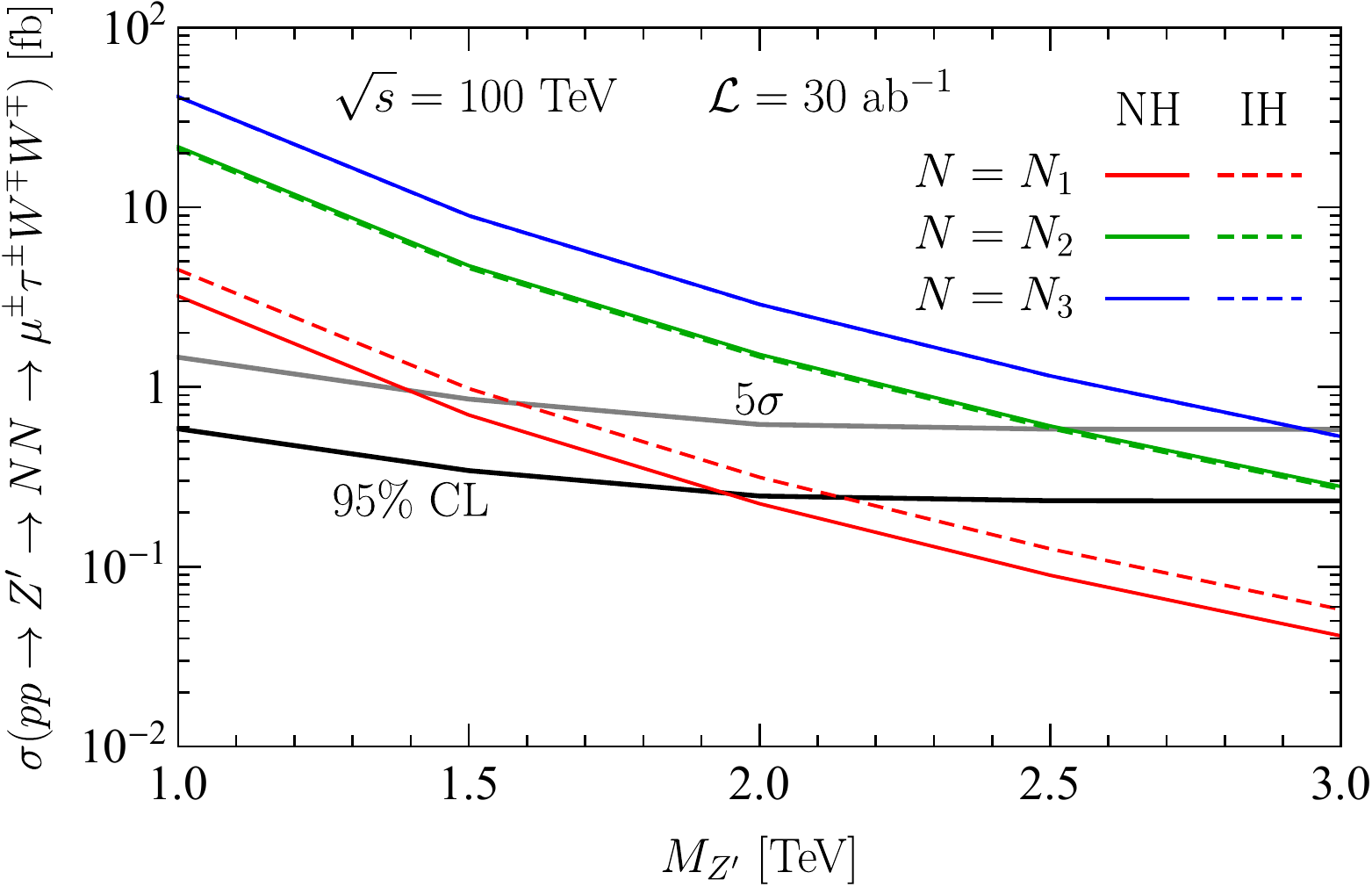}
\caption{Projected sensitivity of cross section $\sigma(pp\to Z'\to NN\to \mu^\pm \tau^\pm W^\mp W^\mp)$ (black and gray curves) at $\sqrt{s}=14$~TeV (top) and $\sqrt{s}=100$~TeV (bottom) for the benchmark model with $M_N = M_{Z'}/4$. The red, green and blue solid (dashed) curves show the predicted cross sections for $N=N_1,N_2,N_3$, respectively, assuming $g'=0.6$ and NH (IH) neutrino mass spectrum.}
\label{fig:sensitivity}
\end{figure}

\begin{figure}[htb!]
\centering
\includegraphics[width=0.48\textwidth]{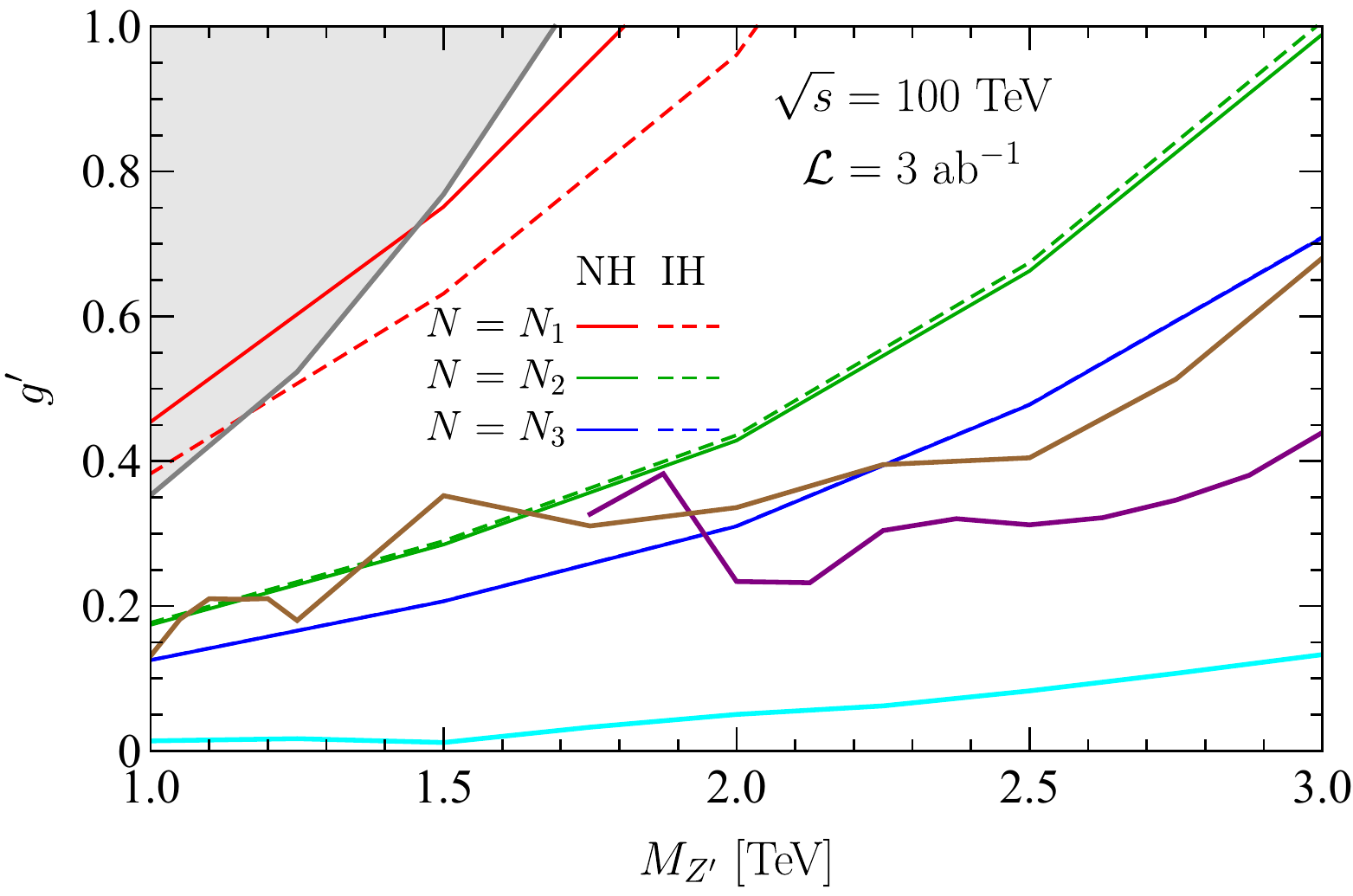}
\includegraphics[width=0.48\textwidth]{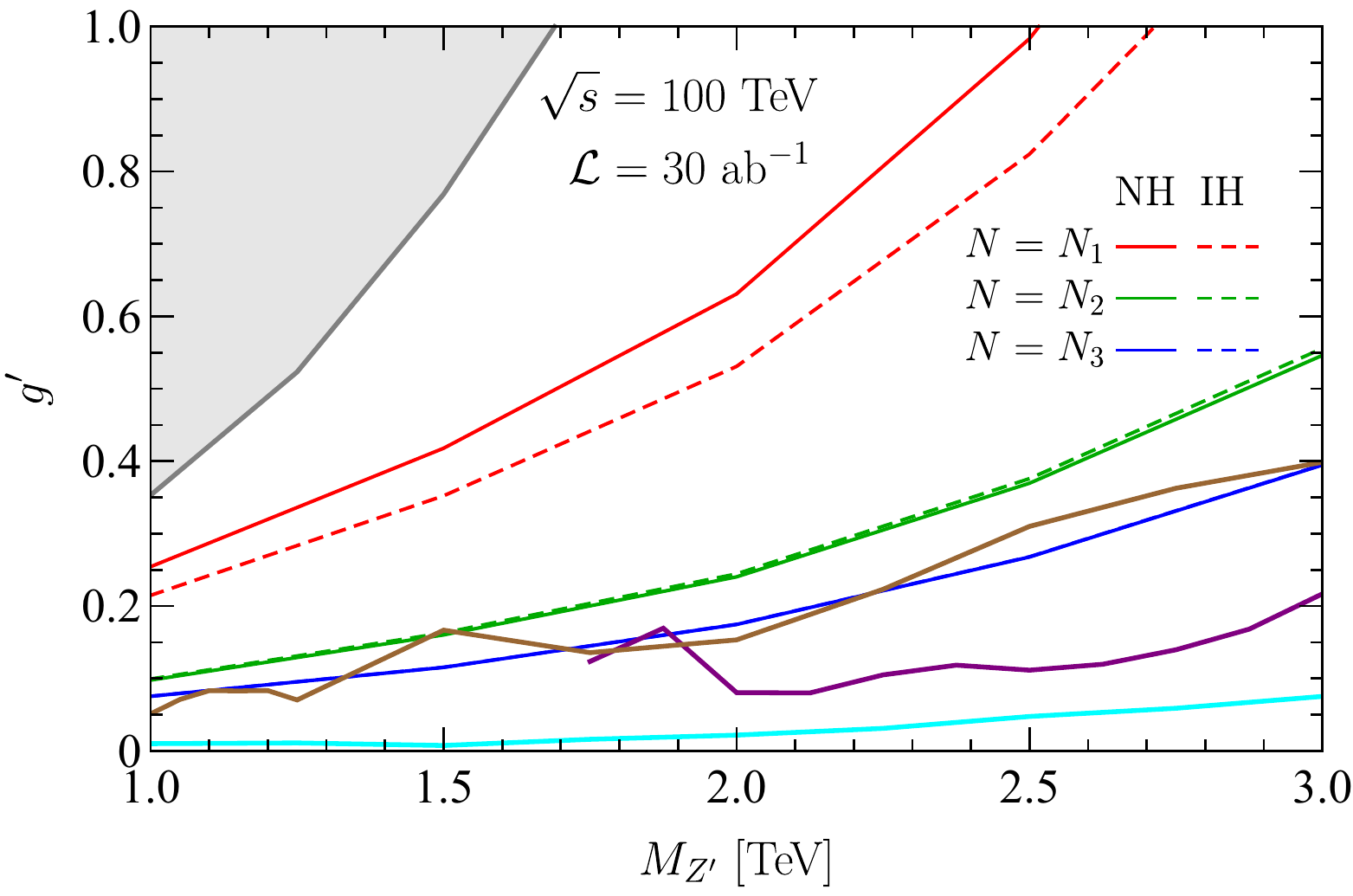}
\caption{Projected sensitivity of the gauge coupling $g^{\prime}$ versus $M_{Z'}$ with the integrated luminosity of 3 ab$^{-1}$ (left) and 30 ab$^{-1}$ (right) at $\sqrt{s}=100$~TeV for the benchmark model with $M_N = M_{Z'}/4$. The red, green and blue solid (dashed) curves correspond to $N=N_1,N_2,N_3$, respectively, assuming NH (IH) neutrino mass spectrum. The gray curve shows the 95\% C.L. limit from the search for $Z^{\prime}\to\tau\tau$ at LHC~\cite{Aaboud:2017sjh}. The estimated exclusions from $Z'\to \tau\tau$ (cyan), $Z'\to b\bar{b}$ (brown) and $Z'\to t\bar{t}$ (purple) at FCC-hh are also shown.
}
\label{fig:g_sensitivity}
\end{figure}

Given the characteristic features of the signal and backgrounds discussed above, we can use the Gaussian method to calculate the significance
\begin{eqnarray}
S/\sqrt{B},
\end{eqnarray}
where $S$ and $B$ are the signal and background event expectations, respectively. With the proper branching fractions in Table~\ref{BR}, the discovered signal events for this channel read as
\begin{eqnarray}
&&S=L\times \sigma_0\times 2\times{\rm BR}^2(W^\pm\to q\bar{q}')\times\epsilon^{\tau}_{NN},
\label{S-typeI}
\end{eqnarray}
where $\sigma_0\equiv\sigma(pp\to Z'\to NN)\times {\rm BR}(N\to \tau^\pm W^\mp)\times {\rm BR}(N\to \ell^\pm W^\mp)$,
$L$ is the integrated luminosity and the factor of 2 accounts for the charge-conjugation of final states. The $\epsilon^{\tau}_{NN}$ denotes the selection efficiency for our signal with one tau lepton. The gray and black curves in Fig.~\ref{fig:sensitivity} (a) respectively correspond to the sensitivities of $\sigma(pp\to Z'\to NN\to \mu^\pm \tau^\pm W^\mp W^\mp)=2\sigma_0$ for 5$\sigma$ discovery and 95\% C.L. exclusion, assuming an expected luminosity of 3 ab$^{-1}$ at 14 TeV LHC and $M_N=M_{Z'}/4$. These sensitivities are applicable to the search for heavy neutrino pair production in any U$(1)'$ extension model. Given the realistic decay branching fractions of $N$ in Table~\ref{BR}, for our U$(1)_{\rm (B-L)_3}$ model, we find that HL-LHC can produce at most 20 LNV events for $M_{Z'}=1$ TeV and cannot reach 5$\sigma$ discovery as shown in Fig.~\ref{fig:sensitivity} (a). We then assess the discovery potential of future circular hadron collider. The recently released Delphes 3.4.2 includes the beta card for FCC studies. Figs.~\ref{fig:sensitivity} (b,c) show the projected sensitivities of the production cross section as a function of $M_{Z'}$, with the integrated luminosity of 3 ab$^{-1}$ (bottom left) and 30 ab$^{-1}$ (bottom right) at $\sqrt{s}=100$~TeV. Compared with the predicted cross sections by our U$(1)_{\rm (B-L)_3}$ model, one can see that the FCC-hh can discover the LNV signal with tau lepton for $M_{Z'}$ up to 2.2 (3) TeV and $N=N_3$ with the fixed $g'=0.6$ and the integrated luminosity of 3 (30) ab$^{-1}$. The projected sensitivities are highly dependent on the benchmark decay branching fractions for each heavy neutrino but not quite on the neutrino mass patterns. The test on the flavor combinations of SM charged leptons would reveal the specific nature of different heavy neutrinos. In Fig.~\ref{fig:g_sensitivity} the 95\% C.L. exclusion obtained using our search methodology is depicted on the parameter space of $g'$ versus $M_{Z'}$ assuming $M_N=M_{Z'}/4$. The 95\% C.L. limit from the search $Z^{\prime}\to\tau\tau$~\cite{Aaboud:2017sjh} is also shown and one can see the gauge coupling $g'$ greater than 0.35 has been excluded for $M_{Z'}=1$ TeV. The searches for $Z'\to b\bar{b}$~\cite{Aaboud:2018tqo} or $Z'\to t\bar{t}$~\cite{Aad:2020kop} do not place severe constraint due to the suppression by the baryon number. The FCC-hh with the integrated luminosity of 3 (30) ab$^{-1}$ can probe the LNV signature with tau lepton for the gauge coupling $g'$ down to 0.12 (0.07) with $M_{Z'}=1$ TeV and $N=N_3$.

In Fig.~\ref{fig:g_sensitivity} we also interpolate the current bounds of $Z'\to \tau^+\tau^-$~\cite{Aaboud:2017sjh}, $Z'\to b\bar{b}$~\cite{Aaboud:2018tqo} and $Z'\to t\bar{t}$~\cite{Aad:2020kop} and estimate their exclusions at future FCC-hh. One can see that the most optimistic channel to directly observe $Z'$ is from the $Z^\prime \rightarrow \tau^+ \tau^-$ channel. However, in the U(1)$_{\rm B-L}$ model, the right-handed neutrinos are needed to cancel the gauge anomalies and the search for LNV signal is important to confirm the existence of the right-handed neutrinos and explain the neutrino masses. Suppose the $Z'$ resonance can be discovered through the di-tau channel in future, the search for the pair production of heavy neutrinos we explored would provide crucial information for the LNV nature and mass pattern of the heavy neutrinos.

\section{Conclusions}
\label{sec:Concl}

Low energy neutrino oscillation data lead to correlation with the flavor structure of the heavy neutrino decays and tau lepton plays manifest role in such correlation. Due to the low tau identification efficiencies, future colliders with high energy and luminosity enables us to investigate and search for heavy neutrino decaying to tau lepton. Moreover, the decoupled lepton number violation in high energy process for low-scale Type I Seesaw encourages us to consider the extension of the canonical Type I Seesaw. We thus examine the lepton flavor signatures with tau lepton at FCC-hh, through lepton number violating processes in the U$(1)_{\rm B-L}$ extended Type I Seesaw. In the U$(1)_{\rm B-L}$ extended Type I Seesaw, we consider a flavored U$(1)_{\rm (B-L)_3}$ model where only the third generation fermions couple to the new gauge boson $Z'$ and thus the constraint on the new gauge interaction is much weak.
We investigate the impact of up-to-date neutrino oscillation results on neutrino mass models and consequently examine the lepton flavor signatures of heavy neutrinos in the Type I Seesaw.

We study the pair production of heavy neutrinos via $Z'$ mediation in the U$(1)_{\rm (B-L)_3}$ model, with one tau lepton in the subsequent decays: $pp\to Z'\to NN\to \ell^\pm\tau^\pm W^\mp W^\mp\to \ell^\pm\tau^\pm + {\rm jets}$.
The $\tau$ lepton's hadronic decay is assumed in the above channels. We take into account the realistic decay branching ratios of heavy neutrinos determined by neutrino oscillation measurements. The general sensitivities of LNV production cross sections are shown for 5$\sigma$ discovery and 95\% C.L. exclusion.
We find that HL-LHC cannot reach 5$\sigma$ discovery for heavy neutrinos in the U$(1)_{\rm (B-L)_3}$ model.
The FCC-hh can discover the LNV signal with tau lepton for $M_{Z'}$ up to 2.2 (3) TeV with $g'=0.6$ and the integrated luminosity of 3 (30) ab$^{-1}$.
The gauge coupling $g'$ down to 0.12 (0.07) can be probed through the LNV signature with tau lepton by the FCC-hh with the integrated luminosity of 3 (30) ab$^{-1}$ and $M_{Z'}=1$ TeV.
The test on the flavor combinations of SM charged leptons would reveal the specific nature of different heavy neutrinos and the three heavy neutrinos in the Casas-Ibarra parametrization with diagonal unity matrix can be distinguished.

\acknowledgments
TL is supported by the National Natural Science Foundation of China (Grant No. 11975129, 12035008) and ``the Fundamental Research Funds for the Central Universities'', Nankai University (Grants No. 63196013). CYY is supported in part by the Grants No.~NSFC-11975130, No.~NSFC-12035008, No.~NSFC-12047533, by the National Key Research and Development Program of China under Grant No.~2017YFA0402200 and the China Post-doctoral Science Foundation under Grant No.~2018M641621. CH acknowledges support from the Sun Yat-Sen University Science Foundation.

\bibliography{refs}

\end{document}